\documentclass[12pt,a4paper]{conference}

\usepackage{fancyhdr}
\usepackage{graphicx,amsmath,amssymb,cite}
\usepackage{multind}
\makeindex{author} \makeindex{subject}
\newcommand{\beq}{\begin{equation}}
\newcommand{\eeq}[1]{\label{#1}\end{equation}}
\newcommand{\eeqn}{\end{equation}}

\newcommand{\beqa}{\begin{eqnarray}}
\newcommand{\eeqa}[1]{\label{#1}\end{eqnarray}}
\newcommand{\eeqan}{\end{eqnarray}}

\let\bar=\overbar

\newcommand{\Dslash}{\not{\hbox{\kern-4pt $D$}}}
\newcommand{\dslash}{\not{\hbox{\kern-2pt $\del$}}}

\newcommand{\msb}{{\bar{\ssstyle M \kern -1pt S}}}

\begin{document}

\Chapter{Mesonic interactions and their contribution to strong phases in flavor physics}{Meson interactions in flavor physics }{B.~El-Bennich \it{et al.}}
\vspace{-6 cm}\includegraphics[width=6 cm]{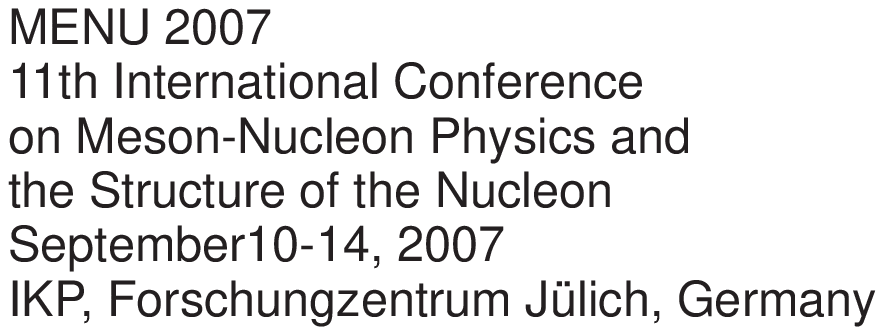}
\vspace{4 cm}

\addcontentsline{toc}{chapter}{{\it N. Author}} \label{authorStart}

\begin{raggedright}


B.~El-Bennich$^{\star}$\footnote{E-mail address: bruno.elbennich@lpnhe.in2p3.fr}~, A.~Furman$^{\dagger}$,
R. Kami\'nski$^{\ddagger}$, L.~Le\'sniak$^{\ddagger}$, B.~Loiseau$^{\star}$ and B. Moussallam$^{\$}$ 

\bigskip\bigskip

$^{\star}$ Laboratoire de Physique Nucl\'eaire et de Hautes \'Energies, \mbox{Groupe Th\'eorie}, 
          Universit\'e Pierre et Marie Curie, 75252 Paris, France \\
$^{\dagger}$  ul. Bronowicka 85/26, 30-091 Krak\'ow, Poland  \\
$^{\ddagger}$ Division of Theoretical Physics, The Henryk Niewodnicza\'nski Institute of Nuclear Physics,
          Polish Academy of Sciences, 31-342 Krak\'ow, Poland    \\
$^{\$}$ Institut de Physique Nucl\'eaire, Universit\'e Paris-Sud,  91406 Orsay, France  \\

\end{raggedright} 

\begin{center}
\textbf{Abstract}
\end{center}

We analyze the contributions of hadronic final-state interactions to the strong phases generated in the $B\to K\pi\pi$ weak decays. To this end, we develop an alternative approach to the commonly employed isobar model based upon scalar and vector form factors for pion-pion and 
pion-kaon interactions.

\section{Preliminary remarks}

An accurate and unequivocal knowledge of strong phases in weak decay amplitudes is crucial to any precision test of $CP$-violating observables. Yet, in heavy meson decays, the decay amplitude is still stricken with hadronic uncertainties. This comprises form factors and subleading contributions such as annihilation amplitudes as well as mesonic final-state interactions.  Here, we concentrate on the latter hadronic contribution; in particular, we investigate the effects of pion-pion and pion-kaon interactions motivated by recent experimental data on $B\to K\pi\pi$ decays~\cite{abe,aubert,garmash,garmash2,aubert2}.  {\em Direct\/} $CP$ violation in $B\to\rho(770)^0 K, \rho(770)^0\to\pi^+\pi^-$ decays was discovered recently~\cite{aubert,garmash}. The three-body decays are commonly analyzed within the isobar model. Several other resonances are observed in the experimental effective $\pi\pi$ and $\pi K$ mass distributions. We point out the  $f_0(980)$ and $\rho(770)^0$, which we recently treated in detail (as well as their interesting interference effects) \cite{elbennich}, and the $K^*(892)$ and $K_0^*(1430)$ which this contribution deals with.

\section{Long-distance form factors in weak decays}

We assume that QCD factorization is applicable in the kinematical configuration in which one pion and the kaon form a quasi colinear pair in the $B$ center of mass frame, where their interaction with the second pion emitted in backward direction is suppressed. Thus, we derive  weak decay amplitudes for a quasi two-body state following  Beneke and Neubert~\cite{bene03}. These amplitudes are given by the product of two factorized currents to which non-factorizable radiative corrections resummable to all orders in $\alpha_s(\mu)$ can be added. The subsequent creation of  a two-pion or a pion-kaon pair in an $S$- or $P$-wave from vacuum is mediated by one of the currents, namely $\langle (\pi\pi)_{S,P}|\gamma_\mu(1- \gamma_5) | 0 \rangle$ or $\langle (\pi K)_{S,P}|\gamma_\mu(1- \gamma_5) | 0 \rangle$, and accordingly described by appropriate scalar or vector form factors which depend on the relative angular momentum and isospin. They can be derived from unitary coupled-channel equations constrained by chiral perturbation theory and experimental data on either $\pi\pi$ or $\pi K$ phase shifts and inelasticities via dispersion relations. 

In the following, we shall concentrate on the case of $(K\pi)_{S,P}$ final-state interactions.  In the $K\pi$ mass range below 2~GeV, the resonances $K^*(892)$ and $K_0^*(1430)$ dominate the pion-kaon vector and scalar form factors, respectively. Here, the $B\to K\pi\pi$ decay amplitudes contain two contributions, one being the QCD factorization amplitudes of weak $b\to s \bar dd$ or $b\to s\bar uu$ transitions previously mentioned, the other a phenomenological long-distance amplitude with either a $c$- or $u$-quark in the loop of the corresponding penguin topology\footnote{~also called charming penguins in the case of long-distance $c$-loops which, close to on-shell, may be associated with intermediate $D_s^{(*)} D^{(*)}$ states.}~\cite{Ciuchini:1997hb}. The $S$-wave part of the $B^-\to (K^-\pi^+)_S\pi^-$ decay amplitude reads
\begin{eqnarray} 
 A_S & = &\frac{G_F}{\sqrt{2}} (M_B^2-m_\pi^2) \frac{m_K^2-m_\pi^2}{q^2}\ F_0^{B\to \pi}(q^2) f_0^{K^-\pi^+}(q^2) \times \nonumber \\
 & \times& \Big \{ \lambda_u(a_4^u + P_u - a_{10}^u/2) +  \lambda_c (a_4^c+P_c-a_{10}^c/2) - \\
 & - & \frac{ 2q^2}{(m_b-m_d)(m_s-m_d)} \big  [ \lambda_u (a_6^u +S_u -a_8^u/2) +\lambda_c (a_6^c+S_c  - a_8^c/2) \big ] \Big \} \nonumber
\end{eqnarray}
while the $P$-wave amplitude is given by
\begin{eqnarray}
  A_P & = & 2 \sqrt{2} G_F \mathbf{p}_{\pi^-}\! \cdot\ \mathbf{p}_{\pi^+} F_1^{B\to \pi}(q^2)f_1^{K^-\pi^+}\!(q^2) \times \nonumber \\
  & \times & [ \lambda_u(a_4^u + P_u - a_{10}^u/2) +  \lambda_c (a_4^c+P_c-a_{10}^c/2) ].
\end{eqnarray}
Here, $q^2$ is the effective $K^-\pi^+$ mass squared, $F_{0,1}^{B\to \pi}(q^2)$ is the scalar/vector  $B\to \pi$ transition form factor and
$f_{0,1}^{K^-\pi^+}(q^2)$ denotes the scalar/vector $K^-\pi^+$ form factor. The $a_i^{u,c}$ are combinations of short-distance Wilson coefficients and the $\lambda_i$ are products of CKM matrix elements (see Ref.~\cite{elbennich} for details). The long-distance penguin amplitudes are parametrized by four complex parameters $S_u,S_c,P_u$ and $P_c$.

\begin{figure}[t!]
\begin{center}
\includegraphics[scale=0.35]{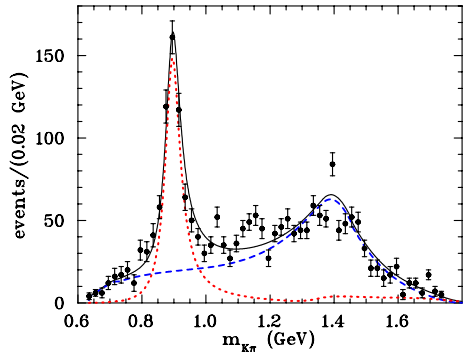}
\caption{Average $K\pi$ effective mass distributions for $B^\pm \to K^\pm \pi^\mp \pi^\pm$ decays. The dashed line represents the $S$-wave, the dotted line denotes the $P$-wave and the solid one corresponds to the total amplitude. The data are from~\cite{garmash}.} 
\label{piKmass}
\end{center}
\vspace*{-6mm}
\end{figure}

\section{Results \& Conclusions}

We are concerned with the four decays $B^\pm \to K^\pm\pi^\mp \pi^\pm$, $\bar B^0 \to \bar K^0\pi^- \pi^+$ and $B^0\to K^0\pi^+\pi^-$ and the corresponding experimental data by the Belle and BaBar collaborations~\cite{aubert,garmash,garmash2,aubert2}. Twelve branching fractions and $CP$-violating asymmetries for $B\to K^*(892)\pi$ and $B\to K^*_0(1430)\pi$ in the above mentioned charge combinations are available as well as 285 data points for the $K\pi$ effective mass and helicity angle distributions. The QCD factorization amplitudes, despite additional strong amplitudes and phases generated by the scalar and vector form factors, do not reproduce the experimental branching fractions. In fact, the theoretical values are too small by a factor spanning from 2.3 to 3.6 if charming penguin amplitudes are not included. Our results then agree with the recent calculations by Cheng, Chua and Soni~\cite{cheng}. We note that annihilation topologies, for which thus far no complete calculation exists, are not accounted for in this work. Including their contributions introduces a parametrization similar to the charming penguin one, though a different scale sub-leading in $\Lambda_{\mathrm{QCD}}/m_b$ is involved.

In Fig.~1, we present some preliminary results on $K\pi$ effective mass distributions and in Fig.~2 their $\pi\pi$  counterpart obtained previously~\cite{elbennich} for comparison's sake. In both cases, charming penguin amplitudes are necessary to reproduce the data. The theoretical curves describe the data structure very well. Furthermore, we can, choosing appropriate $\pi\pi$ and $\pi K$ mass ranges, integrate over these distributions and obtain branching fractions for $B\to f_0(980)K$, $B\to \rho(770)^0 K$, $B\to K^*(892)\pi$ and $B\to K^*_0(1430)\pi$. We stress that our $S$- and $P$-waves are described by single unitary waves that contain all resonant and non-resonant contributions in a unified way. This stands in contrast with the commonly employed isobar model which employs distinct resonant and non-resonant amplitudes. It turns out that a direct comparison with experimentally obtained branching fractions is not straightforward as one has to extract the resonant part of the $K\pi$ $S$- and $P$-waves.

\begin{figure}[t]
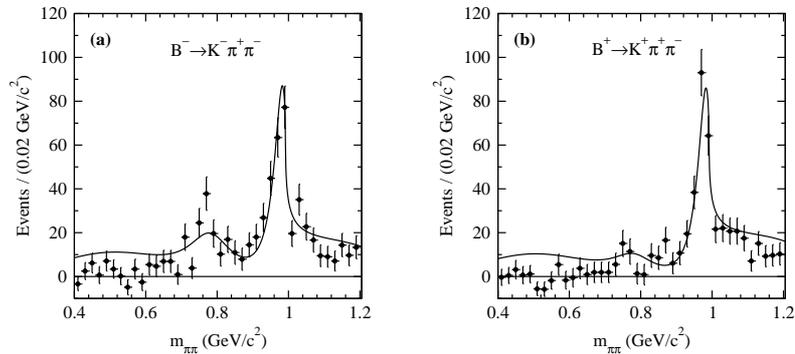

\begin{center}
\includegraphics[scale=0.5]{fig2a.eps}
\hspace{0.5cm}
\includegraphics[scale=0.5]{fig2b.eps}
\caption{$\pi\pi$ effective mass distributions for $B^-\to K^-\pi^+\pi^-$ and $B^+\to K^+\pi^-\pi^+$ decays; the data are taken from~\cite{abe}. Note that $CP$ violation is well visible in the $\rho(770)^0$ mass range.} 
\label{pipimass}
\end{center}
\vspace*{-6mm}
\end{figure}

\section*{Acknowledgments}

This work was supported by the IN2P3-Polish Laboratories Convention (Project No. CSI-12) and the CNRS-Polish Academy of Science agreement (Project No.~19481). B.~E. benefitted from the Marie Curie International Reintegration Grant No. 516228.

\end{document}